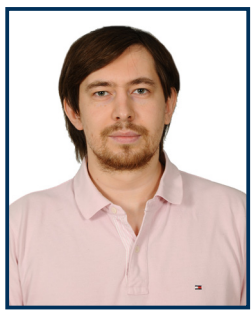 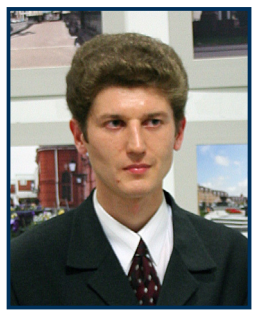

# *A MATHEMATICAL DESIGN AND EVALUATION OF BERNSTEIN-BÉZIER CURVES' SHAPE FEATURES USING THE LAWS OF TECHNICAL AESTHETICS*

*Rushan Ziatdinov*
*Department of Computer and Instructional Technologies,*
*Fatih University, 34500 Buyukcekmece,*
*Istanbul, Turkey E-mail: rushanziatdinov@yandex.ru*
*URL: http://www.ziatdinov-lab.com/*

*Rifkat I. Nabiyev*
*Department of Fine Art and Costume Art,*
*Faculty of Design and National Culture,*
*Ufa State University of Economics and Service, 450068 Ufa, Russia*
*E-mail: dizain55@yandex.ru*

# *МАТЕМАТИЧЕСКИЙ ДИЗАЙН И ОСОБЕННОСТИ ОЦЕНКИ ФОРМООБРАЗУЮЩИХ СВОЙСТВ КРИВЫХ БЕРНШТЕЙНА-БЕЗЬЕ С ПОЗИЦИЙ ЗАКОНОВ ТЕХНИЧЕСКОЙ ЭСТЕТИКИ*

*Рушан Зиатдинов*
*Университет Фатих, г. Стамбул, Турция*
*E-mail: rushanziatdinov@yandex.ru*
*URL: http://www.ziatdinov-lab.com/*

*Р. И. Набиев*
*Уфимский государственный университет экономики*
*и сервиса, г. Уфа, Россия. E-mail: dizain55@yandex.ru*
*URL: http://nabiyev.mathdesign.ru/*

***Abstract.*** We present some notes on the definition of mathematical design as well as on the methods of mathematical modeling which are used in the process of the artistic design of the environment and its components. For the first time in the field of geometric modeling, we perform an aesthetic analysis of planar Bernstein-Bézier curves from the standpoint of the laws of technical aesthetics. The shape features of the curve segments' geometry were evaluated using the following criteria: conciseness-integrity, expressiveness, proportional consistency, compositional balance, structural organization, imagery, rationality, dynamism, scale, flexibility and harmony. In the non-Russian literature, Bernstein-Bézier curves using a monotonic curvature function (i.e., a class A Bézier curve) are considered to be fair (i.e., beautiful) curves, but their aesthetic analysis has never been performed. The aesthetic analysis performed by the authors of this work means that this is no longer the case. To confirm the conclusions of the authors' research, a survey of the «aesthetic appropriateness» of certain Bernstein-Bézier curve segments was conducted among 240 children, aged 14-17. The results of this survey have shown themselves to be in full accordance with the authors' results.

***Keywords:*** MC-curve; spiral; pseudospiral; aesthetic curve; Bernstein-Bézier curves; Bernstein polynomial; monotonicity of curvature; high-quality curve; aesthetic design; spline; computer-aided geometric design; plastic; tension; attraction; structure; aesthetic measure; shape modeling; composition.

***Аннотация.*** В настоящей работе приводятся заметки об определении математического дизайна и методах математического моделирования, находящих применение в процессе художественного конструирования среды и ее компонентов. Впервые в области геометрического моделирования проводится эстетический анализ плоских кривых Берштейна-Безье с позиции законов технической эстетики. Формообразующие признаки геометрии кривых оценивались по следующим критериям: лаконичность-целостность, выразительность, пропорциональная согласованность, композиционное равновесие, структурная организованность, образность, рациональность, динамичность, масштабность, пластичность, гармоничность. В зарубежной литературе кривые Берштейна-Безье с монотонной функцией кривизны (class A Bézier curve) считаются эстетическими кривыми, хотя их эстетический анализ никогда не проводился. Проведенный авторами настоящей работы детальный эстетический анализ показал, что это утверждение ошибочно. Для подтверждения умозаключений авторов было проведено анкетирование среди 240 подростков 14-17 лет на предмет «эстетической целесообразности» тех или иных сегментов кривых Берштейна-Безье, и его результаты показали полное соответствие с результатами авторов.

***Ключевые слова:*** кривая Берштейна-Безье; многочлен Бернштейна; монотонность кривизны; кривая высокого качества; эстетический дизайн; сплайн; компьютерный геометрический дизайн; пластика; напряжение; притяжение; структура; эстетическая оценка; формообразование; композиция.

*Введение*

Математический дизайн — это интегративное научно-художественное направление, инструментарий которого призван выстроить процесс художественного конструирования среды и ее компонентов на основе математического анализа стоящей проблемы с целью получения продукта дизайна, эстетические и технические свойства которого оптимизированы точными расчетами [1-2].

Проще говоря, математический дизайн есть теория математических моделей, применяемых в различных видах дизайна: промышленном, графическом, информационном, архитектурном, ландшафтном и др. Однако, в отличие от чисто математических наук, в математическом дизайне исследуются задачи и проблемы различных видов дизайна на математическом уровне, а результаты представляются в виде теорем, графиков, таблиц и т. д.

Важнейшую роль в математическом дизайне играет математическое моделирование, в частности, следующие его этапы:

• Построение эквивалента объекта дизайна (информационная модель), отражающего в математической форме важнейшие его свойства — законы, которым он подчиняется, структурные связи, присущие составляющим его частям, и т. д. Математическая модель объекта дизайна (или ее фрагментов) исследуется методами теоретической математики, что позволяет получить важные предварительные знания об объекте; математическая модель предоставляет всю необходимую гибкость и творческую свободу в процессе гармонизации формы продукта дизайна;

• Выбор (или разработка) алгоритма для реализации модели на ЭВМ;

• Создание программы, переводящей модель и алгоритм на доступный для ЭВМ язык.

Следует отметить, что данное выше определение в зарубежной литературе отчасти известно под названием Computer Aided Geometric Design [3], что в переводе означает «геометрический дизайн с помощью компьютера». Принимая во внимание тот факт, что изначальные значения терминов дизайн (от лат. designare — отмерять, намечать) и геометрия (от др.-греч. γῆ — Земля и μετρέω — измеряю) по своей сути связаны с измерениями, использование их в виде словосочетания, по всей видимости, в некоторой мере является тавтологией, т.е. в данном случае представляет собой необоснованное повторение близких по смыслу слов. По мнению доктора филологических наук, проф. И.Ш. Юнусова, Computer Aided Geometric Design является неудачно выбранным термином.

*Кривые Бернштейна-Безье*

В 1912 году российский математик Сергей Бернштейн в конструктивном доказательстве аппроксимационной теоремы Вейерштрасса предложил использовать многочлены, которые впоследствии получили название многочленов Бернштейна. Спустя пятьдесят лет они были использованы в качестве базисных функций для параметрических кривых и поверхностей Бернштейна-Безье независимо друг от друга Пьером Безье (Pierre Bézier) из автомобилестроительной компании «Рено» и Полем де Кастельжо (Paul de Faget de Casteljau) из компании «Ситроен», где применялись для проектирования кузовов автомобилей. Со временем, в особенности в капиталистических странах, имя Сергея Натановича Бернштейна потихоньку начало исчезать из названия столь замечательных кривых и сегодня в таких областях, как компьютерная графика, геометрическое моделирование и математический дизайн, они популярны под названием кривых и поверхностей Безье.

В настоящей работе кривые Бернштейна-Безье рассматривается как один из элементов математического дизайна, определение которого было сформулировано выше.

В области геометрического моделирования и промышленного дизайна широкое применение получили кривые с монотонной функцией кривизны (MC-curves) [4-5], которые, в частности, изучались в работах авторов данной работы [6-8]. В [4-5] впервые в области геометрического моделирования был проведен эстетический анализ и оценка структуры и пластических качеств псевдоспиралей с позиций законов технической эстетики.

В зарубежной литературе кривые Бернштейна-Безье с монотонной функцией кривизны (class A Bézier curve) считаются эстетическими кривыми (fair curves) [9], хотя их эстетический анализ никогда не проводился. Проведенный авторами настоящей работы детальный эстетический анализ с позиций законов технической эстетики показал, что это утверждение ошибочно. Ниже приводится его краткое описание.

*Эстетический анализ кривых Бернштейна-Безье*

Проектная практика, осуществляемая в сфере наукоемкого промышленного производства средствами математического моделирования промышленных изделий и их оценкой на участке изготовления материализованного промышленного образца, призвана обеспечить максимальную рациональность, экономичность и эффективность функционирования изделия на всем протяжении его эксплуатации. Однако, эксплуатационные свойства промышленного образца не ограничиваются только техническими характеристиками. Изделие функционирует во всем многообразии связей с человеком, который реагирует на объективные раздражители и оценивает их не только на основе рациональных суждений и умозаключений, но и в аспекте эмоционально-чувственного отношения к миру. В этом смысле уже на этапе предпроектного анализа проблемы в будущий промышленный образец должна закладываться модель не только рациональной, но и его эстетической целесообразности. Подобное диалектическое единство трансформируется в гармонически целостный образ, рождающий побудительный мотив: стимул к эмоциональному восприятию формальных качеств формы как фактору возникновения стремления к раскрытию полезных качеств изделия, что в целом и определит оценочное суждение о нем.

С позиций технической эстетики достижение единства рационального и эмоционального аспектов в образе изделия определяется объективными закономерностями формообразования. В плане их объективизации важно выявить характеристики первоэлементов формы, содержание которых во многом закладывает качественные свойства проектируемого образа изделия.

В качестве такого первоэлемента (геометрической основы) была взята кривая Бернштейна-Безье. Её геометрические свойства анализировались и оценивались с точки зрения их эстетической целесообразности в единстве со способностью отвечать требованиям рациональности. Для этого было выбрано определённое количество математических моделей кривой Бернштейна-Безье с разными геометрическими признаками, формирующими соответствующий визуальный образ кривой. В основу анализа и оценки формообразующих, пластических и выразительных качеств кривых был положен принцип «структурного единства формы». Формообразующие при-

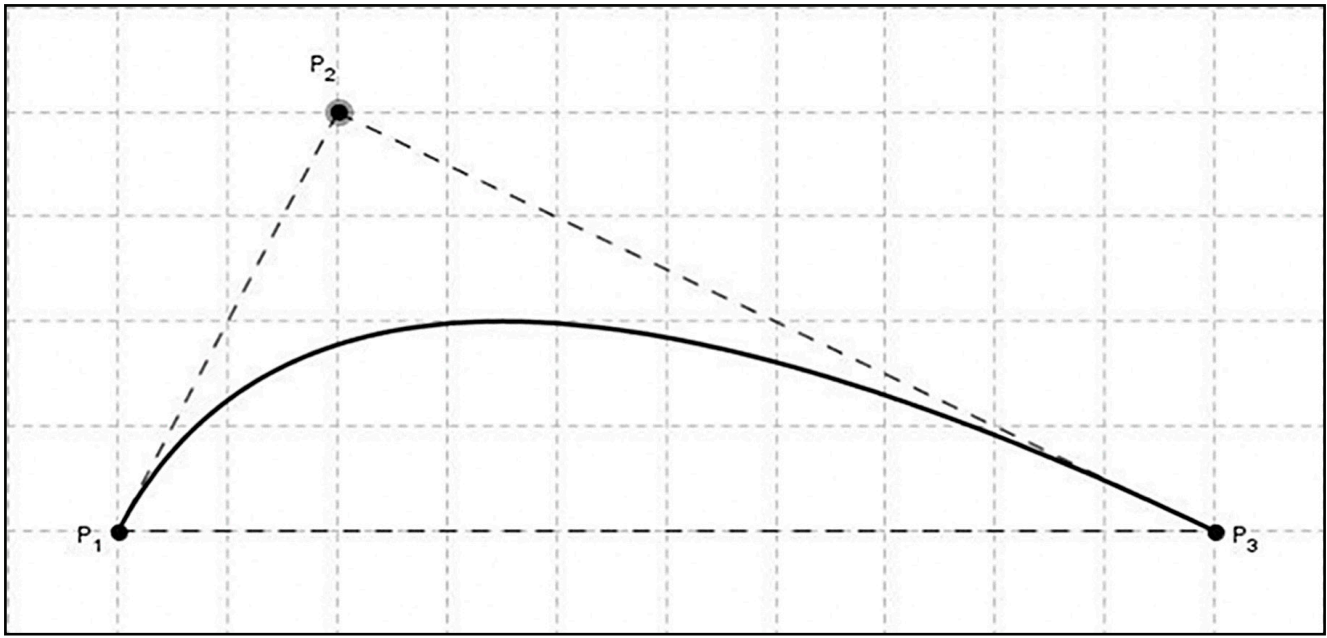

*Рис. 1. Кривая Бернштейна-Безье, функция кривизны которой не монотонная.*

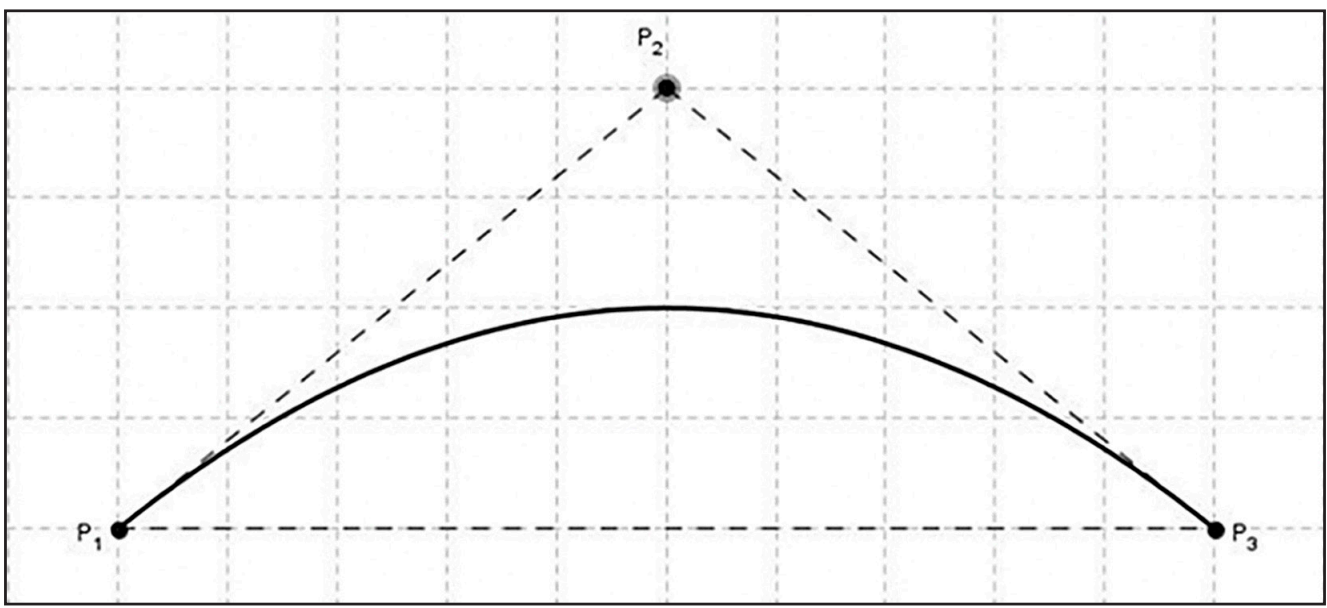

*Рис. 2. Кривая Бернштейна-Безье, функция кривизны которой не монотонная.*

знаки геометрии кривых оценивались по следующим критериям: лаконичность-целостность, выразительность, пропорциональная согласованность, композиционное равновесие, структурная организованность, образность, рациональность, динамичность, масштабность, пластичность, гармоничность.

На основе художественно-конструкторского анализа имеющихся образцов кривых Бернштейна-Безье выявлены закономерности формообразования на уровне геометрических признаков кривой как первоосновы объемно-пространственной организации формы. Важно также подчеркнуть, что объективизация данных осуществлялась методом анкетирования. Его целью было

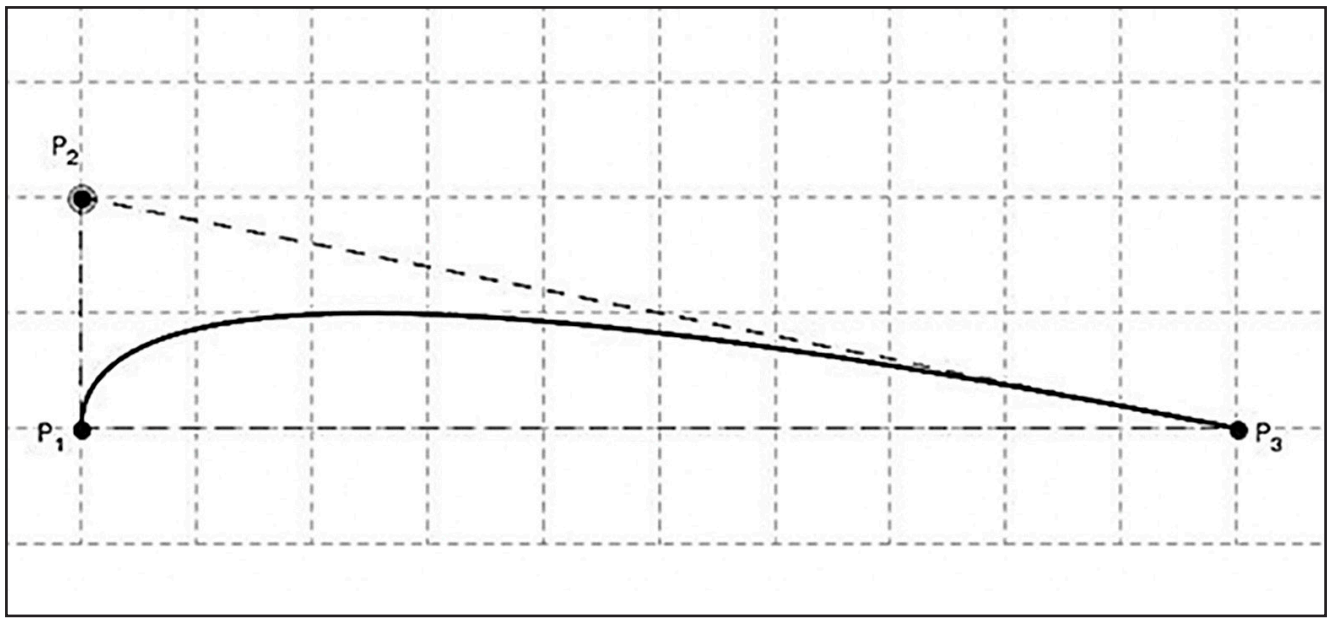

*Рис. 3. Кривая Бернштейна-Безье, функция кривизны которой не монотонная.*

дифференцировать оценку структуры кривых профессионалами, создающих промышленные образцы изделий и их дизайн, а также эмоционально-чувственную реакцию обычных потребителей на характеристики кривых Бернштейна-Безье, также предложенных им для оценки.

Художественно-конструкторский анализ не исчерпывает всех аспектов рассматриваемой проблемы и имеет перспективу дальнейшего развития.

*Заключение*

В рамках настоящего исследования был проведен детальный эстетический анализ 24-х сегментов кривых Бернштейна-Безье второго порядка, 8 из которых имели монотонную функцию кривизны. Оценка по каждому из одиннадцати вышеописанных критериев осуществлялась по семибальной шкале от -3 до 3 (максимальная выраженность, средняя выраженность, минимальная выраженность, осутствие выраженности критерия, минимальное нарушение, среднее нарушение, максимальное нарушение).

Проведенный анализ показал, что у четырех сегментов кривых Бернштейна-Безье с монотонной функции кривизны округленное среднее значение эстетичности (RAVF - rounded average value of fairness) по всем критериям есть 0, что говорит об отсутствии выраженности критериев, у трех сегментов - 1, т.е. критерии выражены минимально, и у одного сегмента кривой выявлено минимальное нарушение критериев.

Для подтверждения умозаключений авторов было проведено анкетирование среди 240 подростков 14-17 лет на предмет «эстетической целесообразности» тех или иных сегментов кривых Бернштейна-Безье и его результаты показали полное соответствие с результатами авторов.

*Благодарность*

Авторы выражают благодарность д. тех. н. профессору Кенджиро Т. Миура (Университет Сидзуока, г. Хамамацу, Япония) и членам его научно-исследовательской лаборатории реалистического моделирования за полезное обсуждение работ по кривым с монотонной функцией кривизны и их применению в геометрическом моделировании и промышленном дизайне.

*Литература*

[1] Набиев Р. И., Зиатдинов Р. А. (2013). Заметки об определении математического дизайна, Материалы XII международной конференции "Системы проектирования, технологической подготовки производства и управления этапами жизненного цикла промышленного продукта (CAD/CAM/PDM-2012)". Институт проблем управления РАН, г. Москва, 2013, стр. 236.

[2] Arslan A., Tari E., Ziatdinov R., Nabiyev R. (2014) Transition Curve Modelling with Kinematical Properties: Research on Log-Aesthetic Curves, Computer Aided Design & Applications., Vol. 11, No. 5,

pp. 508–516.

[3] Farin G. (2001). Curves and Surfaces for CAGD, Morgan Kaufmann, 5th edition.

[4] Ziatdinov R., Nabiyev R., Miura K. T. (2013). MC-curves and aesthetic measurements for pseudospiral curve segments, Mathematical Design & Technical Aesthetics, Vol. 1, No. 1, pp. 6–17.

[5] Зиатдинов Р. А., Набиев Р. И., Миура К. (2013). Т. О некоторых классах плоских кривых с монотонной функцией кривизны, их эстетической оценке и приложениях в промышленном дизайне, Вестник Московского авиационного института, Том 20, № 2. стр. 209–218.

[6] Ziatdinov R., Yoshida N., Kim T. (2012). Analytic parametric equations of log-aesthetic curves in terms of incomplete gamma functions, Computer Aided Geometric Design, Vol. 29. No. 2, pp. 129–140.

[7] Ziatdinov R., Yoshida N., Kim T. (2012). Fitting G2 multispiral transition curve joining two straight lines, Computer-Aided Design, Vol. 44, No. 6, pp. 591–596.

[8] Ziatdinov R. (2012). Family of superspirals with completely monotonic curvature given in terms of Gauss hypergeometric function, Computer Aided Geometric Design, Vol. 29, No. 7, pp. 510—518.

[9] Walton D. J., Meek D. S. (1999). Planar G2 transition between two circles with a fair cubic Bézier curve, Computer-Aided Design, Vol. 31, No. 14, pp. 857–866.